\documentclass[12pt, a4paper]{article}

\baselineskip=\normalbaselineskip

\usepackage{mathrsfs,amsbsy,amssymb,latexsym,amsfonts,amsmath,amsthm}
\usepackage{graphicx,color}

\makeatletter
\catcode`\@=11
\@addtoreset{equation}{section}

\def\@seccntformat#1{\csname the#1\endcsname.~~}
\makeatother

\begin{document}
\begin{titlepage}
\renewcommand{\thefootnote}{\fnsymbol{footnote}}
\begin{flushright}
KUNS-2680
\end{flushright}
\vspace*{1.0cm}

\begin{center}
{\Large \bf 
Distance between configurations\\ 
in Markov chain Monte Carlo simulations
}
\vspace{1.0cm}

\centerline{
{Masafumi Fukuma}%
\footnote{E-mail address: 
fukuma@gauge.scphys.kyoto-u.ac.jp},  
{Nobuyuki Matsumoto}%
\footnote{E-mail address: 
nobu.m@gauge.scphys.kyoto-u.ac.jp} and
{Naoya Umeda}%
\footnote{E-mail address: 
n\_umeda@gauge.scphys.kyoto-u.ac.jp}%
}

\vskip 0.8cm
{\it Department of Physics, Kyoto University, Kyoto 606-8502, Japan}
\vskip 1.2cm 

\end{center}

\begin{abstract}

For a given Markov chain Monte Carlo algorithm 
we introduce a distance between two configurations 
that quantifies the difficulty of transition 
from one configuration to the other configuration. 
We argue that the distance takes a universal form 
for the class of algorithms 
which generate local moves in the configuration space. 
We explicitly calculate the distance for the Langevin algorithm, 
and show that it certainly has desired and expected properties as distance. 
We further show that 
the distance for a multimodal distribution 
gets dramatically reduced from a large value 
by the introduction of a tempering method.
We also argue that, when the original distribution is highly multimodal 
with large number of degenerate vacua,
an anti-de Sitter-like geometry naturally emerges 
in the extended configuration space.

\end{abstract}
\end{titlepage}

\pagestyle{empty}
\pagestyle{plain}

\tableofcontents
\setcounter{footnote}{0}

\section{Introduction}

In Markov chain Monte Carlo (MCMC) simulations, 
one often encounters a situation 
where the equilibrium distribution is multimodal 
and the computation requires an extraordinarily long computer time 
to make the system reach global equilibrium. 
In order to speed up the relaxation to global equilibrium, 
one usually implements an additional method 
such as the overrelaxation \cite{Creutz:1987xi} 
or the simulated/parallel tempering 
\cite{Marinari:1992qd,Swendsen1986,Geyer1991,Earl2005}. 
However, 
it is not always possible to find a nice overrelaxation method. 
Also, the tempering method requires an adjustment of tempering parameter 
such that each local move has a finite amount of acceptance rate, 
which is usually done in a manual or adaptive way 
(and sometimes in an empirical way). 
Thus, it should be useful 
if there is a numerical measure 
that quantifies how much two configurations are separate
for a given MCMC algorithm 
and how fast the system approaches global equilibrium. 
Such measure then can be used 
to numerically evaluate how the situation gets improved 
by the introduction of additional methods, 
which will make the above adjustment easier.

So far there have been proposed various methods for the above purpose, 
but they only quantify the separation 
between two different probability distributions 
(such as the $L_1$ distance) 
or measure the similarity between two data points for a given observable 
(such as the autocorrelation),  
and there has still not been a useful apparatus 
that directly measures an effective distance 
between two different configurations. 
In this paper, 
for a given MCMC algorithm we introduce 
a distance between two configurations 
that quantifies the difficulty of transition 
from one configuration to the other configuration. 
We argue that the distance takes a universal form 
for the class of algorithms 
which generate local moves in the configuration space. 
We make explicit calculations of distance for the Langevin algorithm, 
and show that it has desired and expected properties 
as distance which quantifies the extent of separation 
between two configurations. 
We further show that 
the distance for a multimodal distribution 
gets dramatically reduced from a large value 
by the introduction of a tempering method.

The introduction of such distance opens a way 
to investigate relaxation processes in an MCMC algorithm 
in terms of the geometry of the configuration space itself 
(which should not be confused with the geometry 
of the set of probability distributions). 
As an example, 
we argue that, when the original distribution is highly multimodal 
with large number of degenerate vacua,
our distance can be regarded as a geodesic distance 
with respect to an anti-de Sitter-like (AdS-like) metric. 
Such a geometrical viewpoint enables us 
to adjust the tempering parameter in a purely geometrical way; 
the adjustment can be carried out 
by requiring that the resulting geodesic distance be minimized.

This paper is organized as follows. 
In section \ref{distance}, 
we first introduce a positive, symmetric operator 
(to be called the transfer matrix) 
from the transition matrix of a given MCMC algorithm, 
and then define the distance between two configurations, 
which can be written only with the transfer matrix. 
We argue that the distance takes a universal form 
for the class of MCMC algorithms 
that generate local moves of configuration. 
The statement is confirmed explicitly in Appendix 
for the Langevin and Metropolis algorithms by using a simple model. 
In section \ref{examples}, 
in order to exemplify that the distance actually 
has desired and expected properties,
we study the distance for the Langevin algorithm 
in a one-dimensional configuration space 
with both unimodal and multimodal distributions. 
In section \ref{simtemp}, 
we study the distance for a multimodal distribution 
with the implementation of the simulated tempering method, 
and show that the distance certainly receives 
a huge amount of reduction from a large value. 
We also investigate the local geometry of the extended configuration space 
of the simulated tempering, 
and show that it has an AdS-like geometry. 
Section \ref{conclusion} is devoted to conclusion 
and outlook for future work.

\section{Definition of distance}
\label{distance}

In this section, 
we define the distance between two configurations 
for a given MCMC algorithm. 
The distance will be written 
only with the kernel of a positive, symmetric matrix. 
We argue that the distance should take a universal form 
for the class of MCMC algorithms 
that generate local moves in the configuration space. 

\subsection{Preparation}
\label{distance_preparation}

Let $\mathcal{M}=\{x\}$ be a configuration space, 
and suppose that we are given an MCMC algorithm 
which generates a new configuration $x$ 
from a configuration $y$ 
with the conditional probability $P(x|y)$ at each step. 
We assume that this yields a stochastic process 
which has suitable ergodic properties 
such that $P_n(x|x_0)\equiv (P^n)(x|x_0)$ 
converges to a unique equilibrium distribution $p_{\rm eq}(x)$ 
in the limit $n\to\infty$, 
irrespectively of the initial value $x_0$.
We further make two assumptions: 
\begin{enumerate}
\item 
The algorithm satisfies the detailed balance condition 
for a given real-valued action $S(x)$, 
\begin{align}
 P(x|y)\, e^{-S(y)}
 = P(y|x)\, e^{-S(x)},
\label{detailed_balance}
\end{align}
which ensures that the equilibrium distribution 
is given by $p_{\rm eq}(x)= e^{-S(x)}/Z$ 
($Z=\int dx\,e^{-S(x)}$).
\item
The eigenvalues of $P$ are all positive.%
\footnote{
 Note that the second condition is not too restrictive. 
 In fact, when a transition matrix does not satisfy this condition 
 (i.e., when some of the eigenvalues are negative), 
 one can consider a new transition matrix 
 $P_{\rm new} \equiv (P_{\rm old})^2$, 
 for which the eigenvalues are all positive 
 and the same equilibrium distribution is reached. 
 Note also that this condition is always satisfied 
 for the Langevin algorithm 
 [see, e.g., \eqref{P_Lan}--\eqref{Hamiltonian_Lan2} below].
} 
\end{enumerate}

By using the bra-ket notation 
$P(x|y)=\langle x | \hat{P} | y \rangle$ 
and the configuration operator 
$\hat x \equiv \int dx\,x\,| x\rangle \langle x|$, 
the above two conditions can be rephrased 
as a single statement that 
the operator $\hat{P}\,e^{-S(\hat x)}$ is positive and symmetric. 
This also means that the ``transfer matrix'' 
\begin{align}
 \hat T\equiv e^{S(\hat x)/2}\,\hat{P}\,e^{-S(\hat x)/2}
=e^{S(\hat x)/2}\,\bigl(\hat{P}\,e^{-S(\hat x)}\bigr)\,e^{S(\hat x)/2}
\end{align}
is positive and symmetric.  
$\hat{T}$ shares the same set of eigenvalues as $\hat P$, 
and according to our assumptions, 
all the eigenvalues are positive 
and the largest eigenvalue is unity with no degeneracy. 
Note that $P_n(x_1|x_2)=\langle x_1| \hat{P}^n |x_2\rangle$ 
can be expressed in the form 
\begin{align}
 P_n(x_1|x_2) = K_n(x_1,x_2)\,e^{{}-(1/2)S(x_1)+(1/2)S(x_2)}
\end{align}
with the kernel of $\hat T$,
\begin{align}
 K_n(x_1,x_2)\equiv \langle x_1 |\, \hat T^n | x_2 \rangle.
\end{align}
If we introduce the spectral decomposition of $\hat T$ as%
\footnote{ 
 The ground state wave function is given by 
 $\langle x | 0\rangle = e^{-(1/2)S(x)}/\sqrt{Z}$.
} 
\begin{align}
 \hat T = \sum_{k\geq 0}\lambda_k\,| k \rangle \langle k |
 = | 0 \rangle \langle 0 |
 + \sum_{k\geq 1}\lambda_k\,| k \rangle \langle k |
\end{align}
with $\lambda_0=1 > \lambda_1 \geq  \lambda_2 \geq \cdots > 0$, 
then the kernel is expressed as 
\begin{align}
 K_n(x_1,x_2) = \sum_{k\geq 0} c_k(x_1,x_2)\,\lambda_k^n
\label{kernel_expansion}
\end{align}
with
\begin{align}
 c_k(x_1,x_2) \equiv \langle x_1 | k \rangle\,\langle k | x_2 \rangle. 
\end{align}
The relaxation of the system to global equilibrium 
in the stochastic process 
corresponds to the decoupling of higher modes 
from 
$\hat T^n
 = | 0 \rangle \langle 0 |
 + \sum_{k\geq 1}\lambda_k^n\,| k \rangle \langle k |$ 
as $n$ increases. 
Thus, the large scale behavior of relaxation%
\footnote{ 
 By ``large scale behavior (or structure)'' 
 we mean the behavior (or structure) both at large step numbers 
 and at scales larger than the increment of configuration at each Monte Carlo step.
} 
is determined by the wave functions $\langle x | k \rangle$ 
with small eigenvalues $\lambda_k$.
Note that the decoupling occurs earlier 
for modes with larger $k$.

If we further introduce the Hamiltonian $\hat H$ 
by using a small time increment $\epsilon$ as
$ \hat T = e^{{}-\epsilon \hat H}$,
then $\hat{T}^n$ is expressed as $e^{{}-t \hat H}$ with $t\equiv n\epsilon$.
For generic MCMC algorithms, 
$\hat H$ is a nonlocal operator acting on functions over $\mathcal{M}$. 
However, for the class of MCMC algorithms 
that generate only local moves of configuration, 
$\hat H$ should become local in the limit $\epsilon\to 0$  
(see subsection \ref{distance_universality} 
for further arguments on locality).

We would like to introduce a distance $d_n(x_1|x_2)$ 
for a pair of configurations, $x_1,\,x_2\in\mathcal{M}$, 
such that it enjoys the following properties 
in addition to the usual axioms of distance:
\begin{itemize}
\item
{\bf (P1)}~
The distance $d_n(x_1|x_2)$ vanishes in the limit $n\to\infty$ 
for any pairs $x_1$ and $x_2$, 
in order to reflect the fact 
that any configuration can be reached from every configuration 
in finite steps. 
\item
{\bf (P2)}~
If $x_1$ can be easily reached from $x_2$, 
then the distance is small even for finite $n$. 
\item
{\bf (P3)}~
If the distribution $e^{-S(x)}/Z$ is multimodal, 
and if $x_1$ and $x_2$ belong to different modes, 
then the distance is very large for finite $n$. 
\end{itemize}
As we see below, 
such a distance can be naturally introduced 
if we look at transitions in $\mathcal{M}$ 
that is already in global equilibrium.

\subsection{Half-time overlap}
\label{distance_half-time-overlap}

Let the system be in global equilibrium 
with probability distribution $p_{\rm eq}(x)=e^{-S(x)}/Z$. 
We denote by $\mathcal{X}_n$ the set of sequences of $n$ processes 
in $\mathcal{M}$ 
and by $\mathcal{X}_n(x_1,x_2)$ the subset of $\mathcal{X}_n$ 
that consists of sequences 
which start from $x_2$ and end at $x_1$. 
We then define $f_n (x_1, x_2)$ to be
the ratio of the sizes of two sets: 
\begin{align}
 f_n(x_1,x_2) \equiv \frac{|\mathcal{X}_n(x_1,x_2)|}
 {|\mathcal{X}_n|},
\label{f_n1}
\end{align}
which can be expressed as%
\begin{align}
 f_n(x_1,x_2) = P_n (x_1|x_2)\, \frac{1}{Z}\, e^{-S(x_2)}
 = K_n(x_1,x_2)\, \frac{1}{Z}\, e^{-(1/2)S(x_1)-(1/2)S(x_2)}. 
\label{f_n2}
\end{align}
The latter expression shows that $f_n(x_1,x_2)$ is 
a symmetric function of $x_1$ and $x_2$. 
$f_n(x_1,x_2)$ gives the probability 
to find a sequence of $n$ processes 
from $x_2$ to $x_1$ (or from $x_1$ to $x_2$) 
out of all the possible $n$ processes, 
and thus expresses ``mobility'' between two configurations $x_1$ and $x_2$. 
Note that $f_n(x_1,x_2)$ should take a very small value 
if the equilibrium distribution is multimodal 
and $x_1$ and $x_2$ belong to different modes.

Since our interest is in the mobility 
between two different configurations, 
we normalize the mobility 
such that it takes the maximal value ($=1$) 
for closed loops which start from and end at the same configuration: 
\begin{align}
 F_n(x_1, x_2) &\equiv 
 \frac{f_n(x_1, x_2)}{\sqrt{f_n(x_1, x_1)  f_n(x_2, x_2)}} 
 = \sqrt{\frac{P_n(x_1|x_2)\,P_n(x_2|x_1)}{P_n(x_1|x_1)\,P_n(x_2|x_2)}}.
\end{align}
We will call the normalized mobility $F_n(x_1,x_2)$ 
the {\em half-time overlap} of configurations $x_1$ and $x_2$ 
for $n$ steps. 
In fact, by using \eqref{f_n2} 
and introducing the ``half-time elapsed state'' 
$|x, n/2 \rangle \equiv \hat{T}^{n/2}\,| x \rangle$, 
the function $F_n(x_1,x_2)$ can actually be expressed 
as the overlap of two normalized half-time elapsed states: 
\begin{align}
 F_n(x_1, x_2)
 = \frac{ K_n(x_1,x_2) }
 {\sqrt{ K_n(x_1,x_1)\,K_n(x_2,x_2) }}
 = \frac{ \langle x_1, n/2 \,|\, x_2, n/2 \rangle }
 { ||\, |x_1, n/2\rangle \,|| \, || \,| x_2, n/2\rangle\, ||}. 
\end{align}
One can easily prove that $F_n(x_1,x_2)$ enjoys 
the following properties:%
\begin{align}
 &\bullet~F_n(x_1, x_2) = F_n(x_2, x_1), 
\\
 &\bullet~0 \leq F_n(x_1, x_2) \leq 1, 
\\
 &\bullet~F_n(x, x) = 1, 
\\
 &\bullet~\lim_{n \rightarrow \infty} F_n(x_1, x_2) = 1.
\label{P1a}
\end{align}
Furthermore, from the properties of $f_n(x_1,x_2)$, 
we expect that 
\begin{align}
 &\bullet~\mbox{$F_n(x_1,x_2) \simeq 1$ 
 when $x_1$ can be easily reached from $x_2$ in $n$ steps.}
\label{P2a}
\\
 &\bullet~\mbox{$F_n(x_1,x_2) \ll 1$ 
 when $x_1$ and $x_2$ are separated by high potential barriers.}
\label{P3a}
\end{align}

\subsection{Definition of distance}
\label{distance_definition}

Based on the half-time overlap given above, 
we define a distance between $x_1,\,x_2\in\mathcal{M}$ 
as follows:
\begin{align}
 \theta_n(x_1, x_2) \equiv \arccos(F_n(x_1, x_2)).
\end{align}
This should satisfy the properties (P1)--(P3) 
due to \eqref{P1a}--\eqref{P3a}, 
as well as the following axioms of distance:%
\footnote{
 The axiom of nondegeneracy 
 (i.e., $\theta_n(x_1, x_2) = 0 ~\Rightarrow~ x_1 = x_2$) 
 holds for all the algorithms that involve relaxations. 
 Note that the algorithm consisting of the global updates 
 based on the global heat bath 
 is not in this class, 
 because the system transits to the equilibrium state 
 at the first Monte Carlo step 
 (the transfer matrix is given by the projection, 
 $T=|0\rangle \langle0|$).  
 In fact, for this case,
 the distance $\theta_n(x_1,x_2)$ vanishes 
 for any pairs $x_1$ and $x_2$, 
 which is also consistent with our definition of distance. 
} 
\begin{align}
 &\bullet~\theta_n(x_1, x_2) \geq 0, 
\\
 &\bullet~x_1 = x_2 ~\Leftrightarrow~ \theta_n(x_1, x_2) = 0,
\\
 &\bullet~\theta_n(x_1, x_2) = \theta_n(x_2, x_1),
\\
 &\bullet~\theta_n(x_1, x_2) + \theta_n(x_2, x_3) \geq \theta_n(x_1, x_3).
\end{align}
It is often convenient to introduce other distances 
from the half-time overlap:%
\footnote{
 Note the similarity of the definition of distance 
 between configurations 
 to that between states in quantum information. 
 There, $F_n(x_1,x_2)$ corresponds to the fidelity of two pure states 
 $\rho_{1,2}=| x_{1,2} \rangle \langle x_{1,2} | /
 ||\,| x_{1,2} \rangle \,||^2$, 
 and the distances $\theta_n(x_1,x_2)$ and $D_n(x_1,x_2)$ 
 to Bures length and Bures distance, respectively. 
} 
\begin{align}
 d_n^2(x_1, x_2) &\equiv -2 \log(F_n(x_1, x_2)),
\\
 D_n^2(x_1, x_2) &\equiv 2(1-F_n(x_1, x_2)).
\end{align}
Although these distances do not satisfy the triangle inequality, 
they agree with $\theta_n(x_1,x_2)$ up to higher order corrections 
when $\theta_n(x_1, x_2)$ is small enough. 
For the rest of this paper 
we will mainly use $d_n(x_1,x_2)$ as the definition of distance, 
because it gives the simplest expressions 
for the following examples. 

\subsection{Universality of distance}
\label{distance_universality}

We close this section 
with a comment on the universality of our distance. 
As long as the chosen algorithm generates only local moves of configuration,
we expect that the large scale structure of distance $d_n(x_1,x_2)$ 
takes a universal form, 
in the sense that differences of distance between two such algorithms 
can always be absorbed into a rescaling of $n$. 
In fact, our distance is totally expressed 
with the kernel of the transfer matrix $\hat T$, 
and, when the transition is sufficiently local, 
the corresponding Hamiltonian $\hat H$ is a local operator 
acting on functions over $\mathcal{M}$ in almost the same way. 
We thus expect that 
the eigenvalues $\lambda_k$ and the wave functions $\langle x|k\rangle$
are almost the same for small $k$'s, 
which ensures the same large scale structure of the kernel 
$K_n(x_1,x_2)$
and thus that of the distance $d_n(x_1,x_2)$. 
This statement is explicitly checked in Appendix 
for the Langevin and Metropolis algorithms 
using a simple one-dimensional model. 
Note that the argument for universality are more trustworthy 
when the degrees of freedom of system become larger.

This universality will not hold 
for algorithms that generate nonlocal moves of configuration 
(such as the overrelaxation algorithm).%
\footnote{ 
 The tempering algorithms are nonlocal 
 when viewed from the original configuration space $\mathcal{M}=\{x\}$, 
 but they actually generate local moves 
 in the extended configuration space, 
 so that the universality of the distance is expected to hold 
 also for these algorithms. 
 See arguments below \eqref{metric3} 
 in subsection \ref{simtemp_geometry} for details. 
} 
The Hybrid Monte Carlo (HMC) algorithm \cite{Duane:1987de} 
is marginal in this sense, 
and we leave it for future study 
to investigate 
whether the distance for the HMC algorithm exhibits the same behavior 
as local algorithms.

\section{Examples}
\label{examples}

Expecting the universality of distance commented in the previous section, 
we consider the Langevin method as an MCMC algorithm, 
and write down the explicit form of distance 
for a configuration space $\mathcal{M}=\mathbb{R}$. 
We show that the resulting distance actually satisfies 
the properties (P1)--(P3).

Let $\nu_t$ be the Gaussian white noise 
with diffusion coefficient $D$, 
$\langle \nu_t\, \nu_{t'} \rangle_\nu = 2 D\, \delta(t-t')$, 
and $x_t=x_t(x_0,[\nu])$ be the solution 
to the Langevin equation 
\begin{align}
 \dot{x}_t &= \nu_t - D\,S'(x_t),
 \qquad x_t|_{t=0} = x_0.
\end{align} 
Then, the probability distribution%
\footnote{
 In this section 
 we exclusively use a continuous notation; 
 namely, 
 the distribution $P_n(x|x_0)$ will be written as $P_t(x|x_0)$ 
 with $t=n\epsilon$,
 where $\epsilon$ is the time increment. 
} 
$P_t(x|x_0)\equiv\langle \delta(x-x_t(x_0,[\nu]) \rangle_\nu$ 
is expressed as 
\begin{align}
 P_t(x|x_0) = \langle x | e^{-t \hat{H}_{\rm FP}} | x_0 \rangle
\label{P_Lan}
\end{align}
with the Fokker-Planck Hamiltonian 
(we will set $D=1$ below): 
\begin{align}
 \hat{H}_{\rm FP} ={} -D\,\frac{\partial}{\partial x}\,
 \Bigl[\frac{\partial}{\partial x}+S'(\hat{x})\Bigr].
\end{align}
The transfer matrix is expressed as 
\begin{align}
 \hat T = e^{-\epsilon \hat H},
\label{transfer_Lan}
\end{align}
where the positive, symmetric Hamiltonian $\hat H$ is given by 
\begin{align}
 \hat H &= e^{S(\hat x)/2}\, \hat H_{\rm FP} \, e^{-S(\hat x)/2} 
\nonumber
\\
 &={} \Bigl[{}-\frac{\partial}{\partial x}
 + \frac{1}{2}\,S'(\hat x)\Bigr]
 \Bigl[\frac{\partial}{\partial x}
 + \frac{1}{2}\,S'(\hat x)\Bigr] 
\nonumber
\\
 &={}- \frac{\partial^2}{\partial x^2} + V(\hat x) 
\label{Hamiltonian_Lan}
\end{align}
with
\begin{align}
 V(x) \equiv \frac{1}{4}\,[S'(x)]^2 
 - \frac{1}{2}\,S''(x).
\label{Hamiltonian_Lan2}
\end{align}
The corresponding kernel is then given by 
\begin{align}
 K_t(x,x_0) = \langle x | e^{-t \hat H} | x_0 \rangle
\end{align}
with which the half-time overlap is expressed as 
\begin{align}
 F_t(x_1, x_2) 
 &= \frac{K_t(x_1,x_2)}
 {\sqrt{K_t(x_1,x_1) \, K_t(x_2,x_2)}}. 
\end{align}

\subsection{Example 1: Gaussian}
\label{example_Gaussian}

We first consider the action 
\begin{align}
 S(x) ={} \frac{\omega}{2}\, x^2,
\label{Gaussian}
\end{align}
for which the Hamiltonian $\hat H$ takes the form 
\begin{align}
 \hat H ={} -\frac{\partial^2}{\partial x^2}
 + \frac{\omega^2}{4}\,\hat{x}^2 - \frac{\omega}{2}.
\end{align}
Note that the last term removes the zero-point energy 
of this harmonic oscillator. 
By using the kernel 
\begin{align}
 K_t(x_1,x_2)
 = \sqrt{\frac{\omega}{2\pi\,(1-e^{-2\omega t})}}\,
 \exp\Bigl[{}-\frac{\omega}{4\sinh \omega t}
 \bigl[(x_1^2+x_2^2)\, \cosh \omega t -2 x_1 x_2 \bigr]\Bigr],
\end{align}
the distance $d_t(x_1,x_2)$ is easily obtained to be 
\begin{align}
 d_t^2(x_1, x_2)
 = \frac{\omega}{2 \sinh (\omega t)}\, |x_1 - x_2|^2.
\end{align}
This gives a flat and translation invariant metric 
in the entire configuration space $\mathcal{M}=\mathbb{R}$.%
\footnote{
 Our discussion can be easily generalized to 
 the case $\mathcal{M}=\mathbb{R}^N=\{{\bf x}=(x^i)\}$ and 
 $S({\bf x})=\sum_i \omega_i (x^i)^2$. 
 The distance is then given by
 $d_t^2({\bf x}_1, {\bf x}_2)
 = \sum_i\,(\omega_i/2\sinh (\omega_i t))\,
 |x_1^i - x_2^i|^2.$
} 
Note that the distance decreases exponentially 
as $d_t^2\sim e^{-\omega t}$, 
from which we find that the relaxation time of the system 
is given by $\sim 1/\omega$.%
\footnote{
 If we take the limit $\omega\to 0$ (corresponding to 
 the pure Brownian motion), 
 the distance is given by 
 $d_t^2(x_1, x_2) = (1/2t)\,|x_1 - x_2|^2$, 
 and thus the exponential damping disappears. 
 This reflects the fact that 
 the relaxation time becomes infinite in the limit $\omega\to 0$ 
 in the sense that 
 there is no normalizable equilibrium distribution 
 for the pure Brownian motion. 
} 
Since the manner of relaxation is almost the same for unimodal distributions, 
we see that the distance rapidly decreases to zero 
when the action gives a unimodal distribution. 
We thus confirm that the properties (P1) and (P2) hold in this example.

\subsection{Example 2: perturbation around the Gaussian}
\label{example_perturbation}

We now consider the case 
where the action \eqref{Gaussian} 
is perturbed with a quartic term: 
\begin{align}
 S(x) = \frac{\omega}{2}\, x^2 + \frac{\lambda}{4} x^4.
\end{align}
The perturbative calculation of distance can be done easily, 
and we find to the first order perturbation: 
\begin{align}
 d_t^2(x_1, x_2)
 &=  |x_1 - x_2|^2 \,\Bigl\{ \, \frac{\omega}{2s}
 - \frac{\lambda}{8 \omega s^4}
 \Bigl[ 12 (s^3 - 3s^2c + 3\omega t s + 2 \omega t s^3
 - \omega t s^2c) 
\nonumber
\\
 &~~~+ \omega\, (s^3 + 3s - 3\omega t c) (x_1 - x_2)^2 
\nonumber
\\
 &~~~+ 3 \omega\, (s^3 + 3s - 3\omega t c + 3\omega t -3sc
 + 2\omega t s^2)(x_1 + x_2)^2\Bigr]  + O(\lambda^2)\Bigr\},
\end{align}
where $s \equiv \sinh (\omega t)$ 
and $c \equiv \cosh(\omega t)$. 
This shows that the distance generically does not take 
a flat or translation invariant form 
when the unimodal distribution is not Gaussian.

\subsection{Example 3: double-well potential}
\label{example_double-well}

\begin{figure}[htbp]
\begin{center}
\includegraphics[width=7cm]{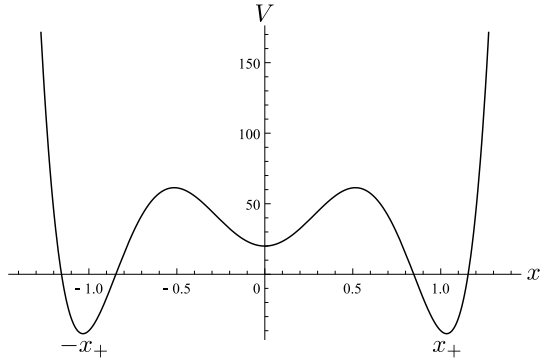}
\begin{quote}
\caption{
The sextic potential $V(x)$ [eq.~\eqref{sextic}] with $\beta=20$, 
which has two global minimum at $x={}\pm x_+$ 
and a local minimum at $x=0$. 
}
\label{figure_LangevinPotential}
\end{quote}
\end{center}
\vspace{-6ex}
\end{figure}

Finally, 
in order to confirm the property (P3), 
we consider the case 
where a high potential barrier exists between local minima: 
\begin{align}
 S(x) = \frac{\beta}{2}\, (x^2-1)^2.  
\end{align}
We assume that $\beta$ takes a large value 
so that the equilibrium distribution $e^{-S(x)}/Z$ is multimodal. 
The potential $V(x)
=(1/4)[S'(x)]^2-(1/2)S''(x)$ becomes sextic 
(see Fig.~\ref{figure_LangevinPotential}),
\begin{align}
 V(x) = \beta^2 x^6 - 2 \beta^2 x^4
 + (\beta^2-3\beta)\,x^2 + \beta,
\label{sextic}
\end{align} 
and has local minima at $x=0$ and $x={}\pm x_+$ 
with $x_+\equiv
\bigl[\bigl(2+\sqrt{1+9/\beta}\bigr)/3\bigr]^{1/2} \simeq 1$. 
One can easily find that $x={}\pm x_+$ are global minima 
by looking at the values of $S''(x)$ 
(note that $S'(x)$ vanishes there).

The eigenvalues $E_k$ $(E_0 = 0 < E_1 < E_2 \cdots)$ 
of $\hat H$ can be roughly estimated as follows. 
We first note that the Gaussian approximation 
around the global minima should be effective 
when $\beta \gg 1$, 
and thus the first two eigenstates of $\hat H$ 
can be approximated as superpositions 
of the ground states $|0_+\rangle$ and $|0_-\rangle$ 
of the approximated Gaussian potential 
$V(x)\simeq (1/2)\,V''(\pm x_+)\,(x\mp x_+)^2
\simeq 4\beta^2\,(x\mp 1)^2$
around the right ($x=x_+\simeq +1$) 
and left ($x=-x_+ \simeq -1$) minima: 
\begin{align}
 |0\rangle &\simeq \frac{1}{\sqrt{2}} 
 \bigl( |0_+\rangle + |0_- \rangle \bigr),
\\
 |1\rangle &\simeq  \frac{1}{\sqrt{2}} 
 \bigl( |0_+\rangle - |0_- \rangle \bigr). 
\end{align}
The energy difference between two states 
is exponentially small, $E_1 = O(e^{-\beta/2})$, 
as can be estimated by an instanton calculation. 

As for the second excitation of $\hat H$, 
we expect the relations 
$E_0=0 \lesssim E_1=O(e^{-\beta/2}) \ll E_2=O(\beta)$.
In fact, there are two possible approximations 
for the second excitation; 
one is to represent the second excitation 
as a superposition of the first excited states 
of the approximated Gaussian potentials 
around the global minima $x={}\pm x_+$, 
and the other is to represent it 
as the ground state of the approximated Gaussian potential 
around the local minimum $x=0$. 
In our case, 
the latter approximation is applicable, 
because the former gives $E_2\simeq 4\beta$ 
and the latter gives $E_2\simeq 2\beta$ 
(see Fig.~\ref{wave_fcns} in Appendix).

By using the expansion of the kernel 
$K_t(x_1,x_2)=\langle x_1| e^{-t H} |x_2 \rangle$
[see \eqref{kernel_expansion}], 
\begin{align}
 K_t(x_1,x_2)
 = \sum_{k\geq 0}\,c_k(x_1,x_2)\,e^{-E_k t}
 \quad
 \bigl( c_k(x_1, x_2) = \langle x_1| k \rangle \,
 \langle k |x_2 \rangle \bigr),
\end{align}
the distance $d_t(x_1,x_2)$ is expressed as 
\begin{align}
 d_t^2(x_1, x_2) &= 
 \sum_{k\geq 1} \frac{(-1)^{k-1}}{k} \Bigl[ 
 \Bigl( \frac{c_1(x_1,x_1)}{c_0(x_1,x_1)}\Bigr)^k 
 + \Bigl( \frac{c_1(x_2,x_2)}{c_0(x_2,x_2)}\Bigr)^k
 -2 \Bigl( \frac{c_1(x_1,x_2)}{c_0(x_1,x_2)}\Bigr)^k
 \Bigr]e^{-k E_1 t} 
\nonumber
\\
 &{}~~~+ \Bigl[ 
 \frac{c_2(x_1,x_1)}{c_0(x_1,x_1)} 
 + \frac{c_2(x_2,x_2)}{c_0(x_2,x_2)} 
 -2 \frac{c_2(x_1,x_2)}{c_0(x_1,x_2)} 
 \Bigr] e^{-E_2 t} + \cdots.
\label{dw_distance}
\end{align}
We now consider the following two cases: 
\begin{align}
  \mbox{Case 1}&: x_1 \simeq x_2 \simeq 1 ,
\nonumber
\\
  \mbox{Case 2}&: x_1 \simeq -x_2 \simeq 1 \nonumber.
\end{align}
We first discuss Case 1 
where $x_1$ and $x_2$ belong to the same mode. 
If we expand $d_t^2(x_1,x_2=x_1+\delta x)$ 
to the second order of $\delta x$, 
the coefficient of $e^{-k E_1 t}$ is given by
\begin{align}
 &\Bigl( \frac{c_1(x_1,x_1)}{c_0(x_1,x_1)}\Bigr)^k 
 + \Bigl( \frac{c_1(x_2,x_2)}{c_0(x_2,x_2)}\Bigr)^k
 -2 \Bigl( \frac{c_1(x_1,x_2)}{c_0(x_1,x_2)}\Bigr)^k
\nonumber
\\
 &\simeq {} 4k^2 \delta x^2\, \biggl(
 \frac{\langle x_1|0_+ \rangle - \langle x_1|0_- \rangle}
 {\langle x_1|0_+ \rangle + \langle x_1|0_- \rangle}\biggr)^{2k}\,
 \biggl(
 \frac{\langle x_1|0_+ \rangle\, \langle x_1|0_- \rangle^\prime
  -\langle x_1|0_+ \rangle^\prime\, \langle x_1|0_- \rangle}
 {\langle x_1|0_+\rangle^2-\langle x_1|0_-\rangle^2}
 \biggr)^2,
\label{coeff_kEit}
\end{align}
which is vanishingly small  
as can be understood by considering the supports of 
the wave functions $\langle x |0_\pm \rangle$. 
Thus, the dominant contribution to the distance 
comes from the second term in \eqref{dw_distance}, 
and the two configurations $x_1$ and $x_2$ 
can be reached from each other 
with a rather short time $\sim 1/E_2=O(1/\beta)$. 
This confirms that two configurations are close to each other 
even for a multimodal distribution 
{\em if} they belong to the same mode. 
In contrast, 
as for Case 2 where $x_1$ and $x_2$ belong to different modes, 
the coefficient of $e^{-E_1 t}$ is not small, 
so that the dominant contribution to the distance 
comes from the first excited state, 
and the transition between $x_1$ and $x_2$ 
requires an exponentially long time $\sim 1/E_1=O(e^{\beta/2})$. 
Accordingly, the distance $d_t(x_1,x_2)$ 
between $x_1\simeq +1$ and $x_2\simeq -1$ 
has a large value 
$d_t^2(x_1,x_2) \propto \beta$,%
\footnote{
 $F_t(x_1,x_2)=e^{-(1/2)\,d_t^2(x_1,x_2)}$ for large $\beta$ 
 can be estimated to be $e^{-{\rm const.} \beta}$ 
 by using an instanton analysis. 
} 
which decreases only very slowly as $t$ elapses. 
We thus see that the property (P3) certainly holds in this example.

\section{Distance for the simulated tempering 
and the emergence of AdS-like geometry}
\label{simtemp}

In this section, 
we show that the value of distance $d_n(x_1,x_2)$ 
for a multimodal probability distribution 
gets dramatically reduced 
when one introduces a tempering algorithm. 
We further argue that 
the effective distance defined for the extended configuration space 
gives an AdS-like geometry 
when the original distribution is highly multimodal 
with large number of degenerate vacua. 

\subsection{Distance for the simulated tempering}
\label{simtemp_distance}

Suppose that we are considering a multimodal distribution, 
and that we implement a simulated tempering method 
\cite{Marinari:1992qd}. 
We take as the tempering parameter 
the overall coefficient $\beta$ of the action 
and introduce the parameter set 
$\mathcal{A}\equiv \{ \beta_a \}_{a = 0,1, \ldots , A}$ 
such that $\beta_0 > \beta_1 > \cdots > \beta_A$, 
where $\beta_0$ is the overall coefficient of the original action. 
We denote by $S(x;\beta_a)$ 
the action with $\beta_0$ replaced by $\beta_a$.

In the simulated tempering algorithm, 
we extend the original configuration space $\mathcal{M}$ 
to $\mathcal{M}\times\mathcal{A}=\{X=(x,\beta_a)\}$, 
and introduce a stochastic process 
such that it converges to global equilibrium 
with the probability distribution 
\begin{align}
 P_{\rm eq}(X) =P_{\rm eq}(x,\beta_a)
 = w_a\, e^{-S(x; \beta_a)}. 
\end{align}
Ideally the weights $w_a$ $(a=0,1,\ldots,A)$ 
are chosen such that the appearance ratio of the $a$-th configuration 
is the same for all $a$ 
[i.e.\ $\int dx\,P_{\rm eq}(x, \beta_a) = 1/(A+1)$].%
\footnote{
 Such weights are given 
 by the inverse of the $a$-th partition functions, 
 $w_a\propto 1/Z(\beta_a)\equiv
 [\int dx\,e^{-S(x;\beta_a)}]^{-1}$. 
 Since it is difficult to evaluate $Z(\beta_a)$ numerically, 
 $w_a$ are usually determined in a manual or adaptive way 
 by investigating the appearance ratio. 
} 
In this paper, we assume that 
$w_a$ are already chosen in this way. 
A possible stochastic process that converges to 
$P_{\rm eq}(X)$ 
is given by a Markov chain 
which consists of the following two steps: 
\begin{itemize}
\item
 \underline{Step 1}.
 Generate a transition in the $x$ direction, 
 $X=(x,\beta_a) \to X'=(x',\beta_a)$, 
 with some proper algorithm 
 (such as the Langevin or Metropolis algorithm).
\item
 \underline{Step 2}.
 Generate a transition in the $\beta$ direction, 
 $X=(x,\beta_a) \to X'=(x,\beta_{a'=a\pm 1})$, 
 with the probability 
 \begin{align}
  \min \Bigl( 1, 
    \frac{w_{a^\prime} e^{-S(x;\beta_{a'})}}
      {w_{a} e^{-S(x;\beta_{a})}}
     \Bigr). 
 \end{align}
\end{itemize}
It is easy to see that each process satisfies 
the detailed balance condition 
with respect to $P_{\rm eq}(X)$. 
After one obtains a sample from the extended configuration space 
with probability distribution $P_{\rm eq}(X)$, 
one estimates the expectation values 
with respect to the original action $S(x;\beta_0)$ 
by using only a subsample with $\beta_{a=0}$.

Since the distribution with smaller $\beta_a$ is less multimodal, 
two configurations belonging to different modes at $\beta_0$
can now be easily reached from each other 
by moving around in the extended configuration space, 
as long as moves in the $\beta$ direction occur frequently 
(as we assume here).  
This improves the convergence to global equilibrium, 
and accordingly, 
the distance between two configurations 
$X_1=(x_1,\beta_0)$ and $X_2=(x_2,\beta_0)$ 
will be reduced. 
To demonstrate that this actually happens, 
we calculated the distance 
between $X_1=(x_1=+1,\beta_0)$ and $X_2=(x_2=-1,\beta_0)$ 
for the action $S(x;\beta_0)=(\beta_0/2)\,(x^2-1)^2$ with $\beta_0=20$, 
by using the simplest setting $A=1$ and $\beta_1=1$. 
As for Step 1 above, 
we adopted the Metropolis algorithm using Gaussian proposal distribution 
with variance $\sigma^2 = 0.01$, 
where the configuration space is restricted to the interval $[-3, 3]$ 
and is latticized with cutoff $a = 0.001$. 
Furthermore, 
by denoting the transition matrix for Step $s$ by $\hat{P}_{(s)}$ $(s=1,2)$, 
we set the transition matrix $\hat{P}$ in the simulated tempering algorithm 
to $\hat{P}=\hat{P}_{(1)} \hat{P}_{(2)} \hat{P}_{(1)}$ 
so that the combined transition matrix 
satisfies the detailed balance condition. 
Below is the result we obtained:%
\footnote{
 In order to align the scale of $n$, 
 the transition matrix $\hat{P}=\hat{P}_{(1)}^{\,2}$ is used 
 for the original local algorithm without tempering 
 (central column in the table). 
}
\\

\begin{tabular}{r||c|c} \hline
 $n$~~ & $d_n^2(X_1,X_2)$ (without tempering)
 & $d_n^2(X_1,X_2)$ (with tempering) \\ \hline \hline
 10 & 39.1 & 26.5 \\ \hline
 50 & 19.2 & 7.16 \\ \hline
 100 & 16.9 & 4.35 \\ \hline
 500 & 13.2 & 0.708 \\ \hline
 1,000 & 11.7 & 0.106 \\ \hline
 5,000 & 8.46 & $2.78 \times 10^{-8}$ \\ \hline
\end{tabular} 
~\\

\noindent
We clearly see that the introduction of the simulated tempering method 
dramatically reduces the distance 
for such a multimodal distribution.

\subsection{AdS-like geometry of the extended configuration space}
\label{simtemp_geometry}

When a system has very large degrees of freedom 
(as in the case of the large-volume or low-temperature limit), 
the issue related to the multimodality of probability distribution 
becomes serious. 
However, it will then become a good approximation 
to coarse-grain the elements of configuration space 
$\mathcal{M}$ 
by regarding configurations in the same mode 
as a single configuration,
and to introduce a new configuration space $\bar{\mathcal{M}}$ 
that consists of coarse-grained configurations 
which are separate from each other. 
We assume that the way of separation between two neighboring configurations 
is uniform over $\bar{\mathcal{M}}$,
keeping in mind a situation 
where the original distribution has highly degenerate vacua.

Suppose that we introduce the simulated tempering algorithm 
to such system. 
The original configuration space $\mathcal{M}$ 
is then extended to $\mathcal{M}\times\mathcal{A}$, 
as discussed in the previous subsection. 
When the degrees of freedom are very large, 
the spacing in the parameter set $\mathcal{A}=\{\beta_a\}$ 
must be sufficiently small 
such that two adjacent configurations 
$(x,\beta_a)$ and $(x,\beta_{a+1})$ 
has an acceptance rate of $O(1)$. 
Then, we may (and we will) regard $\mathcal{A}$ as a continuous set, 
$\mathcal{A}= \{\beta \,|\, \beta_A \leq \beta \leq \beta_0 \}$.

Now let us consider the distance 
in the extended, coarse-grained configuration space: 
$\bar{\mathcal{M}}\times\mathcal{A} = \{X=(x,\beta)\}$. 
We define the metric $ds^2=g_{MN}^{(n)}(X)\,dX^M dX^N$ [$(X^M)=(x,\beta)$] 
such that its geodesic distance between two arbitrarily chosen points 
$X_1=(x_1,\beta_1)$ and $X_2=(x_2,\beta_2)$ 
gives our distance $d_n(X_1,X_2)$. 
\begin{figure}[htbp]
\begin{center}
\includegraphics[width=7cm]{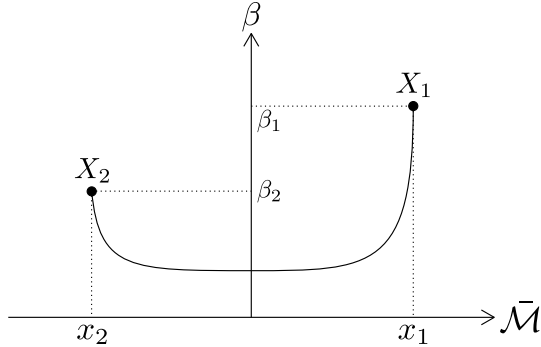}
\begin{quote}
\caption{
The geodesic line for the geometry \eqref{metric} 
with two ends at $X_1=(x_1,\beta_1)$ and $X_2=(x_2,\beta_2)$.}
\label{figure_AdS}
\end{quote}
\end{center}
\vspace{-6ex}
\end{figure}
Such metric will take the following form 
due to the uniformity over $\bar{\mathcal{M}}$:%
\footnote{
 We have assumed that the coefficient of $d\beta\, dx$ is relatively small 
 and can be neglected.
} 
\begin{align}
 ds^2 \simeq f(\beta)\,d\beta^2 + g(\beta)\,dx^2.
\label{metric}
\end{align}
The functions $f(\beta)$ and $g(\beta)$ 
will be studied in detail in the subsequent paper \cite{FMU2}, 
and we here make a simple analysis of their properties. 
Since two different configurations 
$(x_1,\beta)$ and $(x_2,\beta)$ 
in $\bar{\mathcal{M}}\times \mathcal{A}$  
belong to two different modes of the distribution 
$e^{-S(x;\beta)}/Z(\beta)$, 
the transition between them becomes more difficult 
as $\beta$ increases. 
Thus, $g(\beta)$ should be an increasing function at least 
when $\beta$ is large,
and we assume that it takes an asymptotic form $g(\beta)\propto \beta^{q}$ 
for large $\beta$ with some positive number $q$.%
\footnote{
 Without a tempering method, 
 the distance between two configurations belonging to different modes 
 for fixed large $\beta$
 can be roughly estimated by an instanton analysis 
 to be proportional to $\beta$ 
 [see discussions below \eqref{coeff_kEit}]. 
 This implies that the function $g(\beta)$ increases more slowly than $\beta$ 
 when the simulated tempering is implemented, 
 because configurations at larger $\beta$ will get more benefit 
 from the tempering method. 
 We thus expect that $q$ satisfies the inequality $0<q\leq 1$. 
 Our on-going work with numerical calculation \cite{FMU2} 
 shows that $q$ is actually in this range,
 excluding a logarithmic growth of $g(\beta)$. 
} 
On the other hand, 
since the transition in the $\beta$ direction has no obstacle, 
we expect that $f(\beta)$ has a simple behavior. 
Note that, when the measure in the $\beta$ direction is scale invariant 
[i.e., when $f(\beta)\propto 1/\beta^2$] for large $\beta$,
the above metric has the following asymptotic form: 
\begin{align}
 ds^2 \simeq {\rm const.}\, \frac{\rm d\beta^2}{\beta^2}
 + {\rm const.}\, \beta^{q}\,dx^2
 \quad\mbox{($\beta$: large)},
\label{metric2}
\end{align}
which is nothing but the Euclidean AdS metric, 
as can be easily seen by the redefinition of variable, 
$z \equiv {\rm const.}\, \beta^{-q/2}$:
\begin{align}
 ds^2 \propto \frac{1}{z^2}\,(dz^2 + dx^2)
 \quad\mbox{($z$: small)}.
\label{metric3}
\end{align}

We here argue that the appearance of AdS-like geometry 
[eq.~\eqref{metric} with $g(\beta)\sim \beta^{q}$] should be universal. 
In fact, as discussed above, 
when implementing the tempering method 
to an MCMC simulation for a large system, 
the spacing in the parameter set $\mathcal{A}=\{\beta_a\}$ must be 
sufficiently small, 
which allows us to regard $\mathcal{A}$ as a continuous set. 
Then, the stochastic processes in the $x$ and $\beta$ directions 
given in subsection \ref{simtemp_distance}
can be regarded as local moves in the extended configuration space 
$\mathcal{M}\times\mathcal{A}$, 
which in turn define local moves 
in the extended, coarse-grained configuration space 
$\bar{\mathcal{M}}\times\mathcal{A}$. 
Thus, combining with arguments in subsection \ref{distance_universality},
we expect that the distance takes a universal form 
for $\bar{\mathcal{M}}\times\mathcal{A}$ 
in the sense that it does not depend on the particular MCMC algorithms 
that generate local moves along the $x$ and $\beta$ directions.%
\footnote{
 However, the distance should depend on the way of preparing 
 the parameter set $\{\beta_a\}$
 even though the set $\mathcal{A}$ can be regarded as almost continuous. 
} 
This is why we expect that the emergence of AdS-like geometry is universal.

\section{Conclusion and outlook}
\label{conclusion}

In this paper, 
we have defined the distance $d_n(x_1,x_2)$ 
between two configurations $x_1$ and $x_2$ 
for a given MCMC algorithm. 
The distance is written only with the kernel 
of the transfer matrix $\hat T$, 
and various properties of the distance 
(including the universality for local MCMC algorithms) 
are easily understood 
as the reflection of properties of the transfer matrix.

We made a detailed study of the distance 
for multimodal distributions, 
and showed that distances between two different modes 
get dramatically reduced 
by the introduction of the simulated tempering method. 
We also considered the effective distance 
in the extended configuration space 
by regarding configurations in the same mode 
as a single configuration, 
and argued that 
the distance between two configurations 
may then be regarded as the geodesic distance 
in the extended configuration space 
with respect to an AdS-like metric. 
It will be demonstrated in the subsequent paper \cite{FMU2} 
that this is actually the case.

The introduction of distance between configurations opens a way 
to investigate relaxation processes in an MCMC algorithm 
in terms of the geometry of the configuration space itself. 
Among possible applications of the present formalism, 
one interesting application 
is to determine the parameter set $\{\beta_a\}$ 
in the simulated tempering method 
by requiring that it minimize the geodesic distance in the bulk 
with given ends on the boundary at $\beta=\beta_0$. 
Furthermore, 
since the coefficient function $f(\beta)$ in \eqref{metric} 
has a dependence on $\{\beta_a\}$, 
it should be interesting to investigate 
whether the optimized parameter set gives an AdS geometry.  
This point will be argued positively in \cite{FMU2}.

As discussed in subsection \ref{distance_universality}, 
we expect that 
our distance takes a universal form for local MCMC algorithms. 
It should be interesting to study 
to which extent this universality holds. 
In particular, 
it is important to investigate 
whether the HMC algorithm has the distance similar to that for local algorithms.

In this paper, we have discussed only the case where the action $S(x)$ is real. 
When the action takes complex values 
as in QCD at finite density (see \cite{Aarts:2015tyj} for a review), 
we cannot directly use the present definition of distance  
because there can be no real-valued transfer matrix 
for such complex Boltzmann weight. 
The reweighting method gives a real-valued transfer matrix, 
but the reweighting method is generically inapplicable 
because of the sign problem. 
There are various approaches to the sign problem. 
One approach which is currently under intensive study 
is the complex Langevin method \cite{Parisi:1984cs} 
(see also 
\cite{Aarts:2013uxa,Bloch:2017ods,Aarts:2011ax,Hayata:2015lzj,
Nagata:2016vkn,Salcedo:2016kyy}), 
and another is the Lefschetz thimble method \cite{Cristoforetti:2012su} 
(see also 
\cite{Cristoforetti:2013wha,Mukherjee:2013aga,Fujii:2013sra,Cristoforetti:2014gsa,
Alexandru:2015sua,Fukuma:2017fjq,Alexandru:2017oyw,Nishimura:2017vav}).%
\footnote{ 
 Recently a new interesting method has been proposed 
 that uses a complex path 
 optimized with respect to a cost function \cite{Mori:2017pne}.
} 
It must be important to investigate 
whether nice distances can be introduced to these algorithms. 
As for the complex Langevin method, in particular, 
it will be very interesting 
if the condition \cite{Nagata:2016vkn}
to be free from the wrong convergence problems 
\cite{Aarts:2011ax,Hayata:2015lzj,Nagata:2016vkn,Salcedo:2016kyy}
can be rephrased in terms of the distance. 
We expect that the formalism of \cite{Fukuma:2013mx} 
developed for a complex Hamiltonian will be useful.
As for the Lefschetz thimble algorithm, 
it must be interesting 
to investigate the geometry of the tempering algorithm 
that was recently introduced 
for integration over the Lefschetz thimbles, 
where the tempering parameter is set to the flow time 
of the antiholomorphic gradient flow 
\cite{Fukuma:2017fjq,Alexandru:2017oyw}.

A study along these lines is now in progress 
and will be reported elsewhere. 

\section*{Acknowledgments}
The authors thank Yuho Sakatani and Sotaro Sugishita 
for useful discussions. 
This work was partially supported by JSPS KAKENHI 
(Grant Numbers 16K05321 and JP17J08709).

\appendix

\section{Large scale structure of the transfer matrix}
\label{appendix_transfer-matrix}

In this Appendix, 
we explicitly evaluate the transfer matrix 
for the Langevin and Metropolis algorithms 
with a one-dimensional configuration space 
$\mathcal{M}=\mathbb{R}$, 
and show that they have the same large scale structure.

As for the Langevin algorithm, 
if we set the infinitesimal time increment to be $\epsilon$, 
the transfer matrix can be written as 
[see \eqref{transfer_Lan}--\eqref{Hamiltonian_Lan2}]
\begin{align}
 \langle x | \hat T | y \rangle
 =\langle x | e^{-\epsilon \hat{H}} | y \rangle
 \simeq \frac{1}{\sqrt{4\pi\epsilon}}\,
 e^{-(1/4\epsilon)\,(x-y)^2
 -\epsilon\,V((x+y)/2)}
\label{kernel_Lan}
\end{align}
with $V(x)=(1/4)[S'(x)]^2-(1/2)S''(x)$.
As for the Metropolis algorithm, 
if we use a symmetric Gaussian proposal distribution with variance $\sigma^2$, 
the off-diagonal elements of the transfer matrix are given by%
\footnote{
 The diagonal elements can be read off from the condition 
 that $e^{-(1/2)S(x)}$ be the eigenstate of $\hat T$ 
 with unit eigenvalue.
} 
\begin{align}
 \langle x | \hat T | y \rangle
 &=\langle x | \hat{P} | y \rangle\,e^{(1/2)S(x)-(1/2)S(y)}
\nonumber
\\
 &={\rm min}\bigl(1,\,e^{-S(x)+S(y)}\bigr)\,
 \frac{e^{-(1/2\sigma^2)\,(x-y)^2}}{\sqrt{2\pi\sigma^2}}
 \,e^{(1/2)S(x)-(1/2)S(y)}
\nonumber
\\
 & = \frac{1}{\sqrt{2\pi\sigma^2}}\,e^{-(1/2\sigma^2)\,(x-y)^2-(1/2)|S(x)-S(y)|}.
\label{kernel_Met}
\end{align}
We thus see that, 
under the identification $\sigma^2  \sim \epsilon$, 
the Hamiltonians $\hat H ={} -(1/\epsilon)\,\ln \hat T$ 
obtained from \eqref{kernel_Lan} and \eqref{kernel_Met}  
are both local in the limit $\epsilon\to 0$ 
and have the same tendency 
to enhance the values of matrix elements 
when $|x-y|$ and $|S(x)-S(y)|$ are small.

We numerically diagonalize the transfer matrix 
for the action $S(x)=(\beta/2)\,(x^2-1)^2$ with $\beta=20$ 
by latticizing the configuration space. 
Restricting the configuration space to the interval $[-2,\,2]$, 
we set the spatial cutoff $a$, the time increment $\epsilon$ 
and the variance $\sigma^2$ of the proposal distribution in the Metropolis 
to $a=0.005$, $\epsilon=a^2$ and $\sigma^2=2a^2$, respectively. 
Below are the obtained results of eigenvalues:\\ 

 \begin{tabular}{c||c|c||c|c} \hline
  $k$ & $E_k$ (Langevin) & $E_k/E_1$ (Langevin) & 
   $E_k$ (Metropolis) & $E_k/E_1$ (Metropolis) \\ \hline \hline
  0 & 0 & 0 & 0 & 0 \\ \hline
  1 & $7.81 \times 10^{-4}$ & 1 & $7.62 \times 10^{-4}$ & 1 \\ \hline
  2 & 36.2 & $4.63 \times 10^4$ & 34.2 & $4.49 \times 10^4$ \\ \hline
  3 & 58.2 & $7.45 \times 10^4$ & 54.7 & $7.17 \times 10^4$ \\ \hline
 \end{tabular}
~\\

\begin{figure}[htbp]
\begin{center}
\includegraphics[width=15cm]{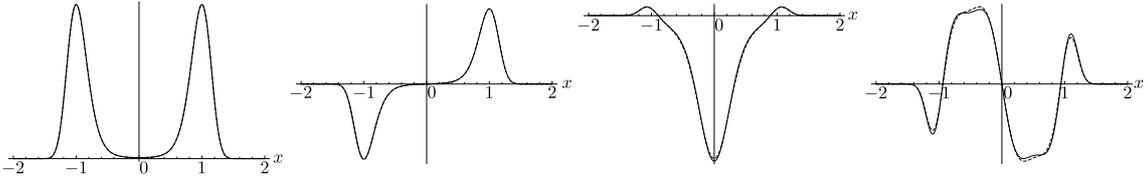}
\begin{quote}
\caption{The eigenfunctions $\langle x | k \rangle$ 
of the transfer matrix 
$\hat T$ ($k=0,1,2,3$ from left to right) 
for the Langevin algorithm (real lines) 
and the Metropolis algorithm (dotted lines). 
We see that they agree with great accuracy. 
}
\label{wave_fcns}
\end{quote}
\end{center}
\end{figure}

\noindent
We see that the eigenvalues for the two algorithms 
can be coincided by rescaling the unit. 
The wave functions also have the same forms 
as depicted in Fig.~\ref{wave_fcns}.

\baselineskip=0.9\normalbaselineskip


\end{document}